\documentstyle[12pt,worldsci]{article}
\newcommand{\ds}{\displaystyle}
\newcommand{\be}{\begin{equation}}
\newcommand{\ee}{\end{equation}}
\newcommand{\vpint}{\ds \int\makebox[0mm][r]{\bf --\hspace*{0.13cm}}}

\newcommand{\trr}{\mathop{Tr}\limits_{x\to y+}}
\begin{document}
\title{QUANTUM MECHANICS OF CONFINEMENT AND CHIRAL SYMMETRY 
BREAKING IN TWO-DIMENSIONAL QCD}
\author{Alexei Nefediev\\
{\em Institute of Theoretical and Experimental Physics, Moscow
117259, Russia}\\}
\maketitle
\setlength{\baselineskip}{2.6ex}

\vspace{0.7cm}
\begin{abstract}

The system of light quark and heavy anti-quark source is studied in 1+1 QCD
in the large $N_C$ limit. Making use of the modified Fock--Schwinger gauge 
allows to consider simultaneously the spectroscopical problem of the 
$q \bar Q$ bound states and the problem of the light quark Green function. 
The Dirac-type equation for the 
spectrum of the system is proved to be equivalent to the well-known 't Hooft
one in the one body limit. 
The unitary transformation from the
Dirac--Pauli representation to the Foldy--Wouthuysen one is carried out 
explicitly, and it is shown that the equation in the Foldy--Wouthuysen 
representation can be treated as a gap equation which defines the 
light quark self-energy in the modified Fock--Schwinger gauge. The 
Foldy--Wouthuysen angle is found to play the role of the Bogoliubov--Valatin 
one and to give the standard value of the chiral condensate.
Connections 
of the given formalism to the standard four-dimensional QCD are outlined and 
discussed.
\end{abstract}
\vspace{0.7cm}
    
    For the first time the two-dimensional model 
of QCD in the limit of infinite number of colours $N_C$
was considered in 1974 by 't Hooft \cite{1} and the celebrated equation of
the same name was derived in the light-cone gauge. 
Four years later, in 1978, this equation was re-derived in the axial gauge \cite{2}.
So the model seems to have been studied well enough. Still it attracts considerable
attention as a problem with features very much similar to those of standard 
four-dimensional QCD. 

Besides the usual assumption $N_C\to\infty$ limit that allows to sum up only planar
diagrammes, we make use of the so-called modified Fock-Schwinger or Balitsky gauge
\cite{3}
\be
A^a_1(x_0,x)=0\quad A^a_0(x_0,0)=0
\label{1}
\ee

As soon as gauge (\ref{1}) is a kind of radial one, the gluonic field can be
expressed in terms of the field strength tensor 
that yields the gluon propagator in the form
\be
K^{ab}_{00}(x_0-y_0,x,y)=
\delta^{ab}\frac{g^2}{2}\delta(x_0-y_0)(|x-y|-|x|-|y|)\equiv 
\delta^{ab}K(x,y),
\ee
and other components equal to zero.

Note that $K$ can be naturally broken into local
$(K^{(1)}\sim |x-y|\delta(x_0-y_0))$ and non-local $(K^{(2)}\sim
(|x|+|y|)\delta(x_0-y_0))$ parts.

Green function for the $q-\bar Q$ system under consideration has the form
\footnote{We adopt the following $\gamma$-matrix convention\cite{2}:
$\gamma_0=\sigma_3$, $\gamma_1=i\sigma_2$, $\gamma_5=\sigma_1$}
$$
S_{q\bar Q}(x,y)=\frac{1}{N_C}\int
D{\psi}D{\bar\psi}DA_{\mu}
\exp{\left\{-\frac14\int 
d^2x
F_{\mu\nu}^{a2} 
-\int
d^2x
{\bar\psi}(i\hat \partial -m -\hat A)\psi \right\}}\times
$$
\be
\times\bar\psi (x) S_{\bar Q} (x,y|A)\psi(y),
\label{7}
\ee
where anti-quark Green function $S_{\bar Q}$ is introduced. The main
advantage of our peculiar gauge choice is the fact that 
the anti-quark is decoupled 
completely so that $S_{\bar Q}$ can be substituted in a very simple form:
\be
\label{8}
S_{\bar Q} (x,y|A)= S_{\bar Q} (x-y);\;
S_{\bar Q}=-i\left(\frac{1+\gamma_0}{2} \theta (t)
e^{-iMt}+\frac{1-\gamma_0}{2}\theta(-t)e^{iMt}
\right)\delta(x).
\ee

On integrating gluon degrees of freedom in (\ref{7}), we arrive at the effective
Lagrangian for the light quark which leads in turn to the 
Schwinger--Dyson equation
\be
(i\hat{\partial}_x-m)S(x,y)+\frac{iN_C}{2}\int
d^2z
\gamma_0S(x,z)\gamma_0K(x,z)S(z,y)=\delta^{(2)}(x-y),
\label{10}
\ee
where
\be
S(x,y)=\frac{1}{N_C}S_{\alpha}^{\alpha}(x,y).
\label{11}
\ee

It is very instructive to note here that the role played by Green function
(\ref{11}) is twofold. By construction $S$ is the Green function of the light 
quark, but due to a very passive part of the static anti-quark it plays the role of
the Green function of the whole $q\bar Q$ system as well, so that the problem of the
light quark Green function and the spectroscopical problem for the $q\bar Q$ system
can be considered simultaneously. We shall get back to this statement later on while
discussing the chiral properties of the model. 

Approach based on spectral decomposition of the Green function (\ref{11}) turns 
out very useful
\footnote{The approach based on diagrammatic technics leads to the same 
results \cite{35}}, so one has
\be
S(q_{10},q_1,q_{20},q_2)=2\pi\delta(q_{10}-q_{20})\left(
\sum_{\varepsilon_n>0}\frac{\varphi_n^{(+)}(q_1)\bar{\varphi}_n^{(+)}(q_2)}
{q_{10}-\varepsilon_n}
+\sum_{\varepsilon_n<0}\frac{\varphi_n^{(-)}(q_1)\bar{\varphi}_n^{(-)}(q_2)}
{q_{10}+\varepsilon_n}\right).
\ee

To proceed further we assume that the Foldy-Wouthuysen operator 
$T(p)=e^{\frac12\theta_F(p)\gamma_1}$
diagonalizing equation (\ref{10}) exists and that angle $\theta_F$ is the same for all
$n$.

With such an assumption applied Schwinger--Dyson equation (\ref{10}) reduces to the
Dirac-type equation in the Hamiltonian form $(
f=\frac{g^2N_C}{4\pi},\alpha=\gamma_0\gamma_1,\beta=\gamma_0,
)$:
\be
(\alpha p +\beta m)\varphi_n^0(p)-
\pi f\int dqdk(\beta {\rm cos}\theta_F(q)+\alpha {\rm sin}
\theta_F(q))K(p-q,k-q)
\varphi_n^0(k)=E_n\varphi_n^0,
\label{13}
\ee
where $\varphi_n^0(p)$ being scalar wave function
\be
\varphi_n^{(+)}(p)=\varphi_n^0(p)T^+(p){1\choose 0},\;
\varphi_n^{(-)}(p)=\varphi_n^0(p)T^+(p){0\choose 1}.
\ee

Let us consider only local ({\it i.e. } generated by $K^{(1)}$) part of interaction 
which reduces to a mass operator $\Sigma$ and 
can be naturally parametrized via two scalar functions $E(p)$ and $\theta(p)$
in the convenient form
\be
\Sigma(p)\equiv\left[E(p)cos\theta(p)-m\right]+\gamma_1\left[E(p)sin\theta(p)-p
\right].
\label{17}
\ee

Self-consistency condition for such a parametrization makes $E(p)$ and $\theta(p)$
satisfy a system of coupled equations \cite{2}:
\be
\left\{
\begin{array}{l}
E(p)cos\theta(p)=m+\frac{\ds f}{\ds 2}\vpint\frac{\ds dk}{(\ds p-k)^2}
cos\theta(k)\\
E(p)sin\theta(p)=p+\frac{\ds f}{\ds 2}\vpint\frac{\ds dk}{\ds (p-k)^2}
sin\theta(k),
\end{array}
\right.
\ee

It is easy to verify that if we identify the Foldy-Wouthuysen angle $\theta_F$ with
the Bogoliubov--Valatin one $\theta$\footnote{$\theta(p)$ obviously plays the role of
the Bogoliubov--Valatin angle as it describes the rotation from the bare particle
with the free dispersion law $\sqrt{p^2+m^2}$ to the dressed \lq\lq physical" 
particle with dispersion $E(p)$.} then the non-local interaction diagonalizes as
well, so that the Foldy-Wouthuysen representation of equation (\ref{13}) takes the
Schr{\"o}dinger-type form
\be
\varepsilon_n\varphi^0_n(p)=E(p)\varphi^0_n(p)-
f\vpint\frac{dk}{(p-k)^2}cos\frac{\theta(p)-\theta(k)}{2}\varphi^0_n(k).
\label{19}
\ee

Equation (\ref{19}) is nothing but the one-body limit of the well-known 't Hooft
equation \cite{2}.

As
mentioned above, Green function of the $q\bar Q$ system constructed from the
solutions of equation (\ref{19}) is the one of the light quark as well, so that
the chiral condensate can be easily calculated
\be
<{\bar q}q>=-i\trr S(x,y)=-\frac{N_C}{\pi}\int_0^\infty dp\; {\rm cos}\theta(p).
\label{20}
\ee

Condensate (\ref{20}) does not vanish in the chiral limit $m\to\infty$ and coincides
with the standard value\cite{6,7}
\be
<\bar{q}q>_{m=0}=-0.29N_C\sqrt{2f}.
\ee

A reasonable question may arise, whether it is worth reproducing old results with a
new complicated method. Still the answer is positive, since the given method can
be easily generalized to describe the four-dimensional QCD. The 
Schwinger--Dyson equation similar to (\ref{10}) takes the following form in Euclidean
space \cite{4}
\be
(-i\hat{\partial}_x-im)S(x,y)+\int d^4z
K_{\mu\nu}(x,z)\gamma_{\mu}S(x,z)\gamma_{\nu}
S(z,y)=\delta^{(4)}(x-y).
\label{22}
\ee

The main object governing the quark dynamics is the correlation function $D$:
\be
<F^a_{\mu\nu}(x)F^b_{\lambda\rho}(y)>=\frac{\delta^{ab}}{N_C^2-1}D(x-y)
(\delta_{\mu\lambda}\delta_{\nu\rho}-\delta_{\mu\rho}\delta_{\nu\lambda})
\ee
with the kernel of equation (\ref{22}) being proportional to $D$.

Function $D$ rapidly decreases at large Euclidean distances and this decrease is 
governed by the gluonic correlation length $T_g$. In the string picture $T_g$ 
defines the radius of the string formed between $q$ and $\bar Q$. 
The two limiting cases, $mT_g\gg 1$ and $mT_g\ll 1$, should be treated separetely,
as it is clearly seen from the non-relativistic expansion of the interaction in 
(\ref{22}).

The case of large $mT_g$ gives quite a natural
result for interaction \cite{5}
\be
V(r)=\left(\frac56+\frac16\gamma_0\right)\sigma r+corrections,
\quad
V_{FW}(r)=\sigma r-\frac{\vec{\sigma}\vec{l}}{4m^2r}+O\left(
\frac{\sigma r}{mT_g}\right),
\label{26}
\ee
which is
in agreement with the Eichten--Feinberg--Gromes results \cite{8,9}, whereas the
opposite limit of small $mT_g$ leads to a self-inconsistency \cite{10}, as the
corrections $\sim O((mT_g)^{-2})$ and diverge in the given limit.

A natural interpretation of such results is offered 
by the string picture of confinement. 
In the case of \lq\lq thick" string $(T_g\gg \frac{1}{m})$ the 
quark interacts
with the gluonic field rather than with a formed string, so the quark 
dynamics is local and potential.

The opposite limit of \lq\lq thin" string  $(T_g\ll \frac{1}{m})$ 
is just the case realized in the two-dimensional
't Hooft model, where strings are infinitely thin (the system just lacks of extra
transverse dimensions to allow the string to swell). The interaction is
sufficiently non-local (see the integral term in the r.h.s. of equation (\ref{19})) and the same behaviour is
expected in the string limit $(T_g\to 0)$ of the four-dimensional QCD.  
Regge picture of the rotating string should follow from
equation (\ref{10}) in this case. Such a picture 
helps to resolve the well-known problem
with local confinement: incorrect trajectory slope for scalar interaction {\it vs}
Klein paradox for vector one.

\end{document}